\documentclass[]{JHEP}
\usepackage{epsfig,cite}
\newcommand{\GeV}{\mbox{~GeV}}

\preprint{Cavendish-HEP-01/09}

\title{The Spectrum of the MSSM with
Non-Standard Supersymmetry Breaking}

\author{J.P.J.~Hetherington\\Cavendish Laboratory, University
of Cambridge, Madingley Road, Cambridge, CB3 0HE, UK}


\abstract{
The Minimal Supersymmetric
Standard Model (MSSM) can include two soft breaking terms which are often
neglected: a non-analytic scalar trilinear coupling
and a Higgsino bilinear term.  
A set of high-scale boundary conditions consistent with the
reparameterisation invariance which the model possesses is obtained.
The three-family renormalisation group equations for the MSSM with these
terms are presented. The ranges of the universal high-scale values of
these couplings which lead to an acceptable TeV-scale theory are obtained,
as is the supersymmetric particle spectrum at
this scale. The effect of the new terms on fine-tuning is presented. 
SOFTSUSY, an existing program for calculating
SUSY particle spectra, has been used, 
with as few modifications as possible.
}

\begin{document}

\section{Introduction}

A possibility for new physics beyond the standard model is supersymmetry
(SUSY)~\cite{martin,haber}. If fermionic generators are added to the bosonic generators of the
Lorentz group, the new space-time symmetry is supersymmetry.
In a theory with exact
supersymmetry, all particles have a partner of equal mass but opposite
spin-statistics. Cancellations between bosonic and fermionic loops prevent
radiative corrections from driving scalar masses up to the highest scale
present,
assumed to be the GUT or Planck scale, $10^{16}$ to $ 10^{19} $~GeV,
solving the naturalness problem of the Standard Model. In addition, the
renormalised electromagnetic, weak, and strong couplings can be made to 
converge to an approximately common
value at the grand unification scale.

Since supersymmetry is not observed 
amongst the already discovered particles it must be a broken
symmetry. The superpartners of the observed particles must be undiscovered
 particles.
The simplest possible SUSY extension of the Standard Model, with a 
new superpartner field for each standard
model field, and the addition of a second Higgs scalar doublet, is called
the minimal supersymmetric Standard Model, or MSSM. 

In the MSSM, the supersymmetry breaking terms are often assumed to be 
flavour-independent and/or gauge-factor
independent at the high scale, and to split as they evolve to low scales under the renormalisation group equations
(RGEs). The model with this
assumption is called the constrained MSSM (CMSSM) or 
minimal supergravity (mSUGRA). The universal nature of the SUSY breaking
terms is motivated by the idea that 
supersymmetry breaking is mediated by
some flavour-blind particle such as the graviton.

Within mSUGRA, the values of fundamental SUSY-breaking parameters are imposed
as boundary conditions on the running
SUSY-breaking masses and couplings at high scale, usually taken to be
$M_X = 10^{16}$~GeV. Experimental masses of superparticles are obtained by evolving the
Lagrangian to the weak scale using the renormalisation
group equations. A solution consistent with the mSUGRA boundary conditions at
$M_X$ and the known Standard Model constraints at $M_Z$ must be found. This
constitutes a boundary value differential equation problem, and can be solved
numerically.

The MSSM Lagrangian is usually claimed to include all possible ``soft supersymmetry
breaking'' terms, terms which split the masses and couplings of particles and their
superpartners, but which do not remove the supersymmetric protection against large
radiative corrections to scalar masses. It is also supposed to include all
possible SUSY conserving terms given the particle content. 

In fact, the ``R-parity violating terms'', which, if unconstrained, allow proton decay
and flavour changing neutral currents, are also possible, but conventionally excluded. If these terms are excluded,
superparticles are created and destroyed only in pairs, giving rise to the usual SUSY phenomenology,
where the lightest supersymmetric particle is stable.

There exists a further set of possible additional 
``non-standard'' soft supersymmetry breaking (NSSB) terms
allowed by the symmetries, as remarked upon in~\cite{martin,Randall,Rosiek,Borz2,Borzumati,JJ}:
a non-analytic scalar trilinear coupling, and a Higgsino bilinear
coupling:
\begin{equation}
{\cal L}^{NSSB} = C^u_{ij}H_1^{*} Q_i \overline{u}_j + C^d_{ij}H_2^{*} Q_i
\overline{d}_j + C^e_{ij}H_2^{*} L_i \overline{e}_j + \tilde{\mu} \tilde{H}_1
\tilde{H}_2 + c.c.,
\label{NSSBterms}
\end{equation}
where $i,j$ are family indices, and weak isospin, spinor, and colour indices
are suppressed. $C^f_{ij}$ and $\tilde{\mu}$ are the new couplings.

In~\cite{Randall} the $C$ terms are written down and then taken as
zero. In~\cite{Rosiek} Feynman rules and mass matrices for these terms are
presented. The 
non-standard terms are written down in~\cite{martin}, and including them in
SUSY studies is advocated there. In~\cite{Borz2} and~\cite{Borzumati} the NSSB terms are used to generate flavour mass hierarchies through radiative corrections. 

In~\cite{JJ} 1-loop $\beta$-functions are obtained for a SUSY model with NSSB
terms, and a general particle content and gauge group. These are then
specialised to a one-family MSSM. The region of 
$m_0, \tan \beta$ parameter space where an
acceptable electroweak vacuum results is obtained, for a given value of
$\tilde{\mu}(M_X)$ and with $\mu=0$. $C_0 =
C^u/Y^u = C^d/Y^d = C^e/Y^e$ is allowed to vary in order to obtain an
acceptable vacuum, as discussed in section~\ref{Highscale} of this
paper. When $C_0 = \tilde{\mu}$ the theory is supersymmetric ($C$ and
$\tilde{\mu}$ can be removed by a reparameterisation as discussed in 
section~\ref{Lagrangian} of this paper.) 
This is a fixed point of the RGEs, and its
stability is discussed in~\cite{JJ}. In~\cite{JJ2} those authors present
2-loop NSSB $\beta$-functions for general particle content and gauge group,
and further discuss the fixed-point properties of the NSSB RGEs. 

In~\cite{Diaz-Cruz:1999fx} the impact of NSSB terms on flavour-changing
neutral currents in the MSSM is considered. In~\cite{Martin:2000hc} the NSSB
terms are included together with dimensionless higher-order scalar couplings
to examine the origin of intermediate scale
physics. In~\cite{Frere:2000uv,Frere:2000ue} the NSSB terms are used in models
of neutrino masses and rare processes in top decay. In~\cite{Ma:2000sq} the
NSSB terms are used to fix the correct mass of the $b$-quark in a model where
the decay $b \rightarrow s \gamma$ is generated radiatively.

These previous works have not considered how the mSUGRA assumptions can be
adapted to be used with NSSB, determined the size of the valid NSSB parameter
space for fixed values of the standard MSSM parameters, 
presented mass-spectra for the NSSB MSSM, or calculated the
fine-tuning of the model. These issues are examined in the present work.

In the following sections the MSSM with NSSB
is
considered in detail.
Section~\ref{problems} is a review of 
the reasons why the NSSB terms have previously
been neglected. In
section~\ref{software} numerical methods which can be used to simulate the
renormalisation group equations are discussed. 
(For more detail on this question,
consult appendix~\ref{numapp}.)  In section~\ref{Lagrangian} a 
reparameterisation invariance is introduced, which the
inclusion of these terms gives to the MSSM. In section~\ref{Highscale}
we develop a version of the ``constrained MSSM'' or ``minimal supergravity''
(mSUGRA) high-scale boundary conditions consistent with the reparameterisation
symmetry. In section~\ref{rges} the calculation of the 3-family
renormalisation group equations (RGEs) for this model is considered. These 
are presented in appendix~\ref{RGEapp}.
The impact of these terms upon electroweak
symmetry breaking is discussed in section~\ref{REWSB}. In
section~\ref{results} an interesting subset of the particle masses
for a standard mSUGRA point but with non-standard breakings is presented. In
section~\ref{constraints} the reasons for the limits on the magnitudes of the
non-standard couplings are explored. In section~\ref{ft} numerical measures
of the 
fine-tuning of the model are considered, and results for the fine-tuning
presented and discussed. Appendix~\ref{RGEapp}
lists the NSSB contribution to the 1-loop renormalisation group equations for
the model. Appendix~\ref{numapp} examines the numerical methods used in this
work.

\section{Why have the NSSB terms traditionally been excluded?}
\label{problems}

The terms of eq.~(\ref{NSSBterms}) have traditionally been neglected because they are thought to lead to large
radiative corrections when there are scalar fields present which are uncharged under the entire gauge
group. 

This argument cannot be used in the 
context of the MSSM, which contains no singlets at the SUSY scale.
In a model with some higher mass
singlet field of mass $m_s$, together with a very massive field
of mass $m_X$, the correction, (see figure ~\ref{fdiag}) to the
relation between the Higgs pole $m_{h,p}$ and
running $m_{h,r}$ masses is:

\FIGURE{
\hbox{\epsfysize=5cm
\epsffile{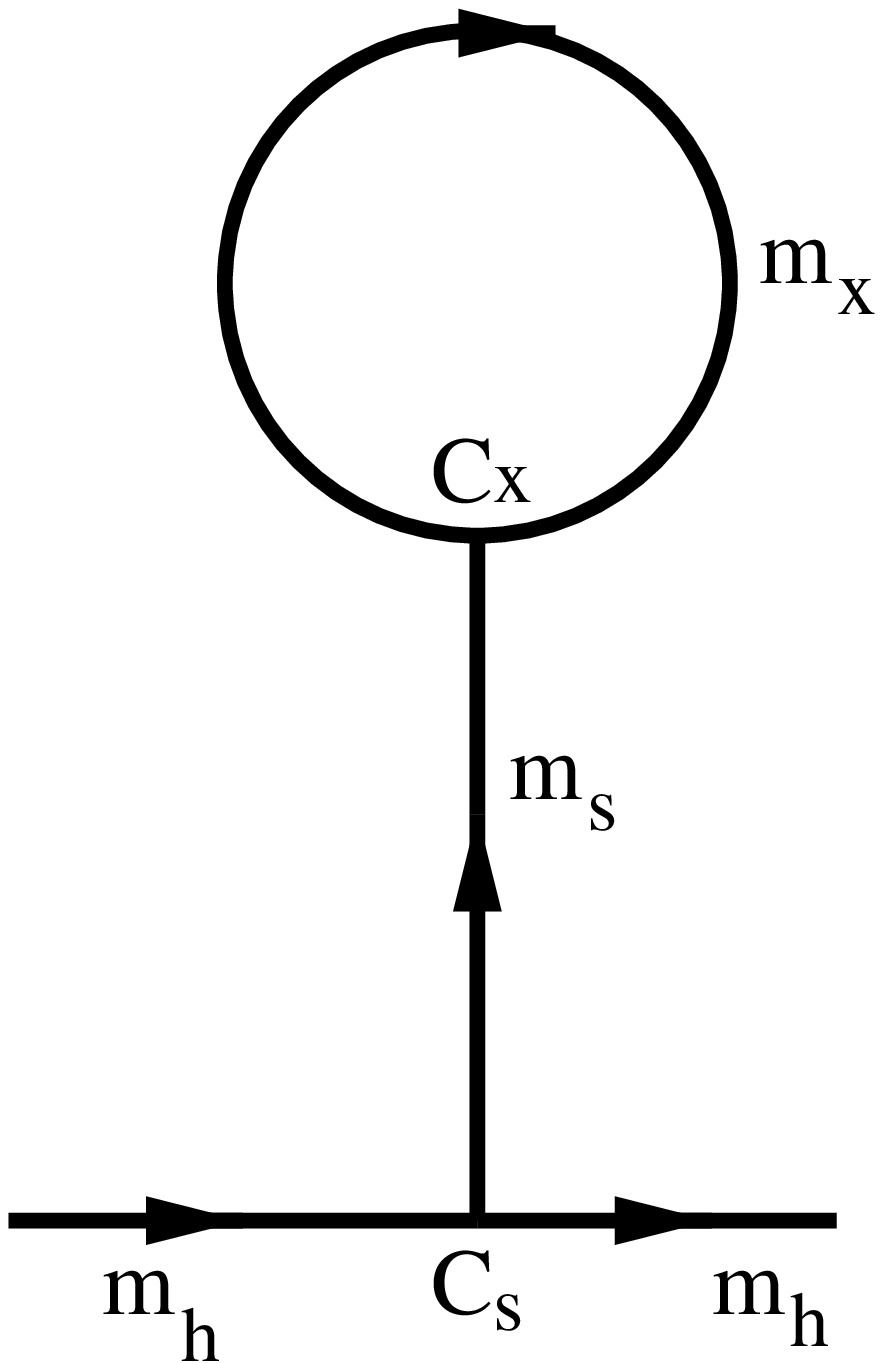}}
\caption{The scalar singlet tadpole correction to the Higgs propagator.
}
\label{fdiag}
}
\begin{equation}
m^2_{h,p} = m^2_{h,r}(m_{h,p}) + C_{s} C_{X} S_4 \frac{m_X^2}{m_s^2}
 \ln(\frac{m_X^2}{m_{h,p}^2}) + ...,
\end{equation}
where $C_s$ is the coupling between the Higgs and the singlet, and $C_X$ is
the coupling between the singlet and the heavy field. $S_4$ is a collection
of dimensionless constants.
If the singlet field has a mass significantly below the high scale, the
theory is thus unstable. However, if 
any singlet fields have masses of order $M_X$, such as in
SUSY desert scenarios, then there is no problem with including NSSB
terms.

\section{Numerical calculation of SUSY particle spectra}
\label{software}

In SUSY models, the 
supersymmetric Lagrangian terms are fixed at the experimental 
scale $M_Z$ from Standard Model particle data. 
The supersymmetry breaking terms are fixed at some high scale, $M_X$,
according to a scheme such as mSUGRA.

The high- and low- scale boundary constraints, together with the
renormalisation group equations, and the intermediate scale radiative
electroweak symmetry breaking (REWSB) constraint (see section~\ref{REWSB}),
form a system of differential equations; a ``boundary value problem''.

The numerical methods used to implement this procedure are discussed in
detail in appendix~\ref{numapp}. Three such methods are discussed. The first
is the formal
``shooting method'' where the constraint equation at one boundary is solved
for those parameters which are unconstrained at the other boundary. This method cannot be
used for SUSY, because of the complex form of the boundary conditions at the
low scale. 

The second method, which may be called the ``drift method''
involves alternate imposition of each set of boundary
conditions. This constitutes a recurrence relation for successive
values of those parameters unconstrained at one of the boundaries. The 
solution consistent with both sets of boundary conditions is a fixed point of
the recurrence relation, but this point is 
not guaranteed to be stable or unique. If we assume the point is both stable
and unique, however, simple repeated imposition of the boundary conditions,
alternately at $M_Z$ and $M_X$, will
result in convergence toward the required solution. This is the method used in
ISASUGRA~\cite{ISASUSY}, and in SOFTSUSY~\cite{softsusy}, the code which was
modified for this work. 
The C++ inheritance structure of SOFTSUSY permitted large parts of the code
of SOFTSUSY to be re-used.

Because the drift method becomes unstable for certain values of the NSSB
parameters, a third approach has been developed, a ``single-ended'' shooting
method, where the fixed point of
the recurrence relation is solved for, using a Newton-Raphson method.
This is further discussed in appendix B.

\section{The NSSB Lagrangian and $\mu$-reparameterisation}
\label{Lagrangian}

The nonstandard SUSY breaking terms to be added to the MSSM Lagrangian are
given in eq.~(\ref{NSSBterms}). Using the usual formula \cite{barger} for extracting the full Lagrangian from the superpotential, the following terms are those involving $m^2_{H_{1/2}}$, the Higgs soft masses, $\mu$, the Higgs bilinear superpotential term, and the NSSB couplings:
\begin{eqnarray}
{\cal L} & = ... & + (\tilde{\mu} - \mu) H_1 H_2 + (\mu^2 +
m^2_{H_{1}}) | H_1 |^2 + (\mu^2 + m^2_{H_{2}}) | H_2 |^2  \nonumber \\
&& + (C^u_{ij} - Y^u_{ij} \mu) H_2^{*} Q_i \overline{u}_j + (C^d_{ij} -
Y^d_{ij} \mu) H_2^{*} Q_i \overline{d}_j + (C^e_{ij} - Y^e_{ij}
\mu) H_2^{*} L_i \overline{e}_j \nonumber \\
&& + c.c.,
\label{lagran}
\end{eqnarray}
where $Y^{u/d/e}_{ij}$ are the Yukawa couplings.

Therefore, the Lagrangian is invariant under the following reparameterisation:
\begin{eqnarray}
\mu & \rightarrow & \mu + \delta \nonumber; \\
\tilde{\mu} & \rightarrow & \tilde{\mu} + \delta \nonumber;\\
m_H^2 & \rightarrow & m_H^2 - 2 \mu \delta - \delta^2 \nonumber; \\
C^f_{ij} & \rightarrow & C^f_{ij} + Y^f_{ij} \delta.
\end{eqnarray}
This was partially pointed out in \cite{JJ} where it was noted that one could
write an MSSM Lagrangian $(\tilde{\mu} = 0, C^f_{ij}= 0, \mu \neq 0)$ in
terms of an NSSB model with zero $\mu$ but non-zero $C^f_{ij}$ and
$\tilde{\mu}$. Here a different emphasis is used:, the Lagrangian contains 
all terms allowed by the symmetries, and the reparameterisation is used 
like a gauge freedom, 
to allow calculations to be made in the most convenient ``gauge''.

The above Lagrangian contains all terms involving $\mu$, provided the Higgs bilinear breaking term is written as 
\begin{equation}
{\cal L} = ... + m_3^2 H_1 H_2
\end{equation}
 and not
\begin{equation}
{\cal L} = ... - B \mu H_1 H_2
\label{oldB}
\end{equation} 
as is sometimes used, factoring out the $\mu$. In the latter case, $B$ would become a $\mu$-reparameterisation dependent quantity. In order that this work can easily make contact with MSSM work conducted using $B$, we instead have the freedom to define $B$ using the Lagrangian term:

\begin{equation}
{\cal L} = ... - B (\mu - \tilde{\mu}) H_1 H_2.
\end{equation} 
On the MSSM subspace of parameter space, where $\tilde{\mu}$ and $C^f_{ij}$
are zero, this definition is consistent with eq.~(\ref{oldB}), 
but $B$ remains $\mu$-reparameterisation invariant. 

\section{The high-scale boundary conditions}
\label{Highscale}

As mentioned in section~\ref{software} the 
supersymmetric Lagrangian terms are fixed at the experimental 
scale $M_Z$ from Standard Model particle data. 
The supersymmetry breaking terms are fixed at some high scale, $M_X$,
according to a scheme such as mSUGRA.

The exceptions to this are the superpotential $\mu$-term, and the SUSY
breaking term $B$. The magnitude of the Higgs vacuum expectation value (VEV) $v^2 = v_1^2 + v_2^2$,
is known
from the Z boson mass. In addition, a choice for $\tan \beta$ at $M_Z$ is
usually made. 
The values of $\mu$ and $B$ fix both $v$ and $v_1/v_2 = \tan \beta$, the
ratio of the two Higgs vacuum expectation values, by minimisation of the
Higgs effective potential (see section~\ref{REWSB})
which is a known function of the model parameters.
Thus, the values of $\mu$ and $B$ cannot be imposed as a
boundary condition, but are allowed to vary to fit the Higgs VEVs.

While the above approach is the conventional one for MSSM calculations, 
in \cite{JJ} the authors chose to use the reparameterisation freedom to take
$\mu=0$, solving the SUSY $\mu$-problem.  
Instead of leaving $\mu$ and $B$ at $M_X$ unconstrained, so that they can
vary to fit  $M_Z$ and $\tan
\beta$, in this approach $C$ and $B$ are left unconstrained at $M_X$.
(With $C^e, C^d, C^u$ taken as equal at
$M_X$ and $\tilde{\mu}$ at $M_X$ given as an extra boundary condition.)

In this investigation, we want to impose values for both $C$ and $\tilde{\mu}$ 
as boundary conditions at $M_X$,
with the same kind of universality assumptions as in mSUGRA,
i.e. $C^{f}_{ij} = Y^{f}_{ij} C_0$. The parameter $\mu$ will
remain in the model, and will, together with $B$, be varied to fit $\tan
\beta$ and $v$. Since we do not choose $\mu = 0$ the model retains its
reparameterisation symmetry.

This approach is perhaps preferable 
because it implements non-standard 
supersymmetry breaking with \emph{minimal} 
changes to mSUGRA. However, there is a subtle difficulty, because we 
wish to use the mSUGRA assumption of universal high-scale soft scalar
masses: the mass term for each of the sleptons, squarks and Higgs bosons
should be the same at $M_X$. 
This assumption is inconsistent with the
reparameterisation invariance. A
reparameterisation changes the Higgs soft masses but not the squark and
slepton masses, breaking the equality.
To make the boundary condition well defined, we must therefore 
select the reparameterisation in which we want the scalars to be universal.

As will be seen in section~\ref{rges}, the renormalisation group running will
take place in $\mu=0$ reparameterisation. If we simply impose the
mSUGRA boundary conditions in that reparameterisation, including a value for
$\tilde{\mu}$, we cannot have $\mu$ as an unconstrained parameter at high
scale.  

Consider a model with $m_0$, $M_{1/2}$, $C_0$ and $\tilde{\mu}$ at
$M_X$ given as constraints,
and values for $\mu$ ($\mu_S$) and $B$ ($B_S$)
and for the other parameters unconstrained at
high scale, but which
are consistent with the low energy constraints. (This discussion is
illustrated in
table~\ref{highscale}). On changing to the $\mu =
0$ reparameterisation used to run the RGEs, or the $\tilde{\mu}=0$
reparameterisation used for some parts of the calculation, the values of $\mu$
and $\tilde{\mu}$ are mixed up, and only $(\tilde{\mu}
- \mu)$ is known. In addition, the Higgs soft mass-squareds become different
from those of the squarks and sleptons.

If we then run the RGEs, do the rest of one cycle of the
calculation, and then return to $M_X$, we do not know how to separate $\mu$
and $\tilde{\mu}$. It seems we cannot return to the reparameterisation used to
impose the constraints. 
However, the difference between the Higgs scalar masses and the squark and
slepton masses must be that caused by the reparameterisation. Thus,
the reparameterisation which makes the squark/slepton soft masses equal to the
Higgs soft masses must be that which returns us to the initial
state. This can be called the ``universality reparameterisation''. Re-imposition of the high-scale boundary conditions then leaves the
model unchanged.

As the calculation converges toward a solution consistent with both sets of
boundary conditions, we need an estimator for the difference between the
Higgs and squark/slepton soft mass-squareds at $M_X$. Until convergence is reached, 
$m_{H_1}
\neq m_{H_2}$, $m_{\tilde{u}} \neq m_{\tilde{d}}$ etc. The choice used here
is the difference
between the average Higgs mass-squared and the average squark mass-squared.

Unlike the $\mu = 0$ approach this approach allows us, for any point in the
MSSM, to model the same point, by using the NSSB with $C_0 = \tilde{\mu} =
0$. The resulting solution is identical with that obtained using SOFTSUSY
normally, and the high-scale value of $\mu$
which the model selects to fit $\tan \beta$ is the same as in the MSSM. 
It is possible to gradually turn on $\tilde{\mu}$ and $C_0$ and move away from any 
MSSM point. 
The MSSM is a $D-2$ dimensional subset of the $D$ dimensional NSSB parameter space.

\TABULAR{|c|c|c|c|c||c|}{
\hline
$\overline{m_{\tilde{q}}}$ & $\overline{m_h^2}$ & $\tilde{\mu}$ & $\mu$ & $C^f
$ & Action \\
\hline
\hline
\multicolumn{6}{|c|}{MSSM as MSSM} \\
\hline
 $m_0$ & $m_0^2$ & N/A & $\mu_c$ & N/A
  & Arrive at high scale \\
\hline
 $m_0$ & $m_0^2$ & N/A & $\mu_c$ & N/A
 & Impose b.c.: at \\
&&&&&convergence this does nothing.\\
\hline
\hline
\multicolumn{6}{|c|}{True NSSB} \\
\hline
 $m_0$ & $m_0^2 + \mu_c^2$  & $-\mu_c$ +
$\tilde{\mu}_c$ & 0 & $(C_0 - \mu_c) Y^f$ & Arrive at high scale in \\
&&&&&$\mu = 0$ reparameterisation \\
\hline
 $m_0$ & $m_0^2$ & $\tilde{\mu}_c$ & $\mu_c$ &
$C_0 Y^f$
 & Reparameterise \\
&&&&&$\delta = \sqrt{\overline{m_h^2} -
\overline{m^2_{\tilde{q}}}} = \mu_c$ \\
\hline
 $m_0$ & $m_0^2$ & $\tilde{\mu}_c$ & $\mu_c$ & $C_0 Y^f$
 & Apply b.c. \\
\hline
 $m_0$ & $m_0^2 + \mu_c^2$ & $-\mu_c + \tilde{\mu}_c$ & 0 & $(C_0 - \mu_c) Y^f$
 & Reparameterise $\delta = -\mu_c$\\
\hline
\hline
\multicolumn{6}{|c|}{MSSM as NSSB} \\
\hline
 $m_0$ & $m_0^2 + \mu_c^2$  & $-\mu_c$ & 0 & $-\mu_c
Y^f$ & Arrive at high scale in \\
&&&&&$\mu = 0$ reparameterisation \\
\hline
 $m_0$ & $m_0^2$ & 0 & $\mu_c$ &
0
 & Reparameterise \\
&&&&&$\delta = \sqrt{\overline{m_h^2} -
\overline{m^2_{\tilde{q}}}} = \mu_c$ \\
\hline
 $m_0$ & $m_0^2$ & 0 & $\mu_c$ & 0
 & Apply b.c. \\
\hline
 $m_0$ & $m_0^2 + \mu_c^2$ & $-\mu_c$ & 0 & $-\mu_c Y^f$
 & Reparameterise $\delta = -\mu_c$\\

\hline
}
{
The high-scale boundary condition(b.c.) part of the iterative procedure for
determining the model parameters consistent with the chosen input parameters.
The state of the model after
convergence is achieved is shown, i.e., once the model has become invariant
under the iterative numerical procedure used. Thus, the imposition of the
high-scale boundary conditions leaves the parameters unchanged, as shown in
the table. \label{highscale}
$\overline{m^2_{\tilde{q}}}$ is the average squark mass-squared,
$\overline{m_h^2}$ is the average Higgs mass-squared. $X_c$ is used to
denote a particular value of
the parameter named $X$.
}

\section{NSSB $\beta$-functions}
\label{rges}

Ref.~\cite{JJ} presents general one-loop beta functions for a general SUSY
model with arbitrary particle content and gauge group, allowing NSSB
terms. The two-loop results are presented in \cite{JJ2}. Reference \cite{JJ}
also gives the RGEs specialised to the NSSB MSSM with one family and
with the Yukawa couplings factored out of the parameters. For example,
the Lagrangian ``$A$'' soft breaking term is written as $m_{10} Y^t H_2 Q \overline{t}$.

An alternative presentation of the beta functions, better suited to numerical
implementation, is the 3-family matrix form for the MSSM without NSSB
used in \cite{barger}, with the
Yukawa matrices not factored out of the soft breakings. This version of the
beta-functions has been calculated.

A subtle question is the impact of the $\mu$-reparameterisation on the beta
functions. The results of \cite{JJ} are in $\mu=0$ reparameterisation. The
MSSM results of \cite{barger} have non-zero $\mu$. 
We do not wish to
re-calculate all of the MSSM part of the beta functions for the $\mu = 0$
reparameterisation. We wish to use part of the results of \cite{barger},
modified according to the general results of \cite{JJ}, which are for $\mu =
0$.
By considering the $\mu$-reparameterisation behaviour of each quantity, and representing the beta functions as pieces with different reparameterisation behaviours, and with dependence on each parameter, it is possible to obtain for the soft masses:
\begin{equation}
\beta_{m_i^2}^{NSSB} = \beta_{m_i^2}^{MSSM} + \beta_{m_i^2}^{C,\tilde{\mu}},
\end{equation}
where the left hand side of each equation is the NSSB beta function for $\mu
= 0$, $\beta^{MSSM}$ is the MSSM $\mu \ne 0$ beta-function \emph{with $\mu$
set to zero}, and $\beta^{C,\tilde{\mu}}$ is the piece from the general
result of \cite{JJ} which has $\tilde{\mu}$ or $C$ dependence. 
A similar equation can be obtained for $\beta_B$, but it is simpler just to calculate it afresh. The beta functions for $C$ and $\tilde{\mu}$ can be obtained from the general expression of \cite{JJ}.

The renormalisation group equations are presented in
appendix~\ref{RGEapp}.

\section{Radiative electroweak symmetry breaking and other low-scale considerations}
\label{REWSB}

The constraints given by analytical 
minimisation of the Higgs effective potential are
used to obtain $\mu$ and $B$ from $\tan \beta$ and $v$. This is done at the
scale ($M_S$) where radiative corrections to the renormalisation group improved
scalar potential have the smallest scale dependence, see below.

The form of the tree-level radiative electroweak symmetry breaking (REWSB)
 constraint equation for $\mu$:

\begin{equation}
\frac{1}{2} M_Z^2 = \frac{m_{H_1}^2 + \mu^2 - (m_{H_2}^2 + \mu^2) \tan^2 \beta}{\tan^2 \beta - 1}
\label{REWSBcon}
\end{equation}
suggests that the resulting value of $\mu$ will be correct in any reparameterisation,
because of the way in which the reparameterisation keeps $m_{H}^2 + \mu^2$
constant. This equation finds the $\mu$ value consistent with given Higgs soft
masses. However, in the $\mu = 0$ reparameterisation, numerical implementation of
this equation fails, because the resulting $\mu^2$ value close to convergence
toward zero is sometimes negative.

The values of the model parameters obtained after imposing the REWSB
constraints, if compared in the same reparameterisation, must be the same
whatever reparameterisation we impose eq.~(\ref{REWSBcon}) in. This is a consequence
of the invariance of the Lagrangian, and can be verified by considering the
application of a reparameterisation to the tree-level result eq.~(\ref{REWSBcon}). In order that the REWSB routine be
convergent around a positive value of $\mu^2$, we can thus choose to impose
the REWSB constraint after reparameterising so that
$\tilde{\mu}=0$.

The constraint equation~(\ref{REWSBcon}) is obtained by minimisation of the
 effective potential. In our study, such a constraint equation is used to
 obtain $\mu$, and a similar expression to obtain $B$. The one-loop
 corrections to the effective potential are included, with 
the leading top/stop contributions from the
 NSSB couplings, and the calculation is performed at the renormalisation scale
 where the scale dependence of the 1-loop corrections is smallest,
 approximately $\sqrt{m_{\tilde{t_1}}m_{\tilde{t_2}}}$. This
 scale, $M_S$ is approximately 760~GeV for the mSUGRA parameters used here,
and is intermediate between $M_Z$ and $M_X$. It is believed that the scale at
 which the 1-loop result has the slowest scale dependence, the contribution
 from higher loops is smallest. This is motivated by ref.~\cite{gamberini}, 
which
 shows that for the MSSM Higgs potential, the tree-level and 1-loop results
 are closest at this scale. However, this relationship has been shown not to 
work as well
 for other potentials, such as MSSM charge-and-color-breaking potentials, in
 ref.~\cite{ferreira}.

The physical mass matrices for those particles affected by $\tilde{\mu}$ and
$C$ in $\mu=0$ reparameterisation are given in \cite{JJ}. The unaffected mass
matrices can be used as originally implemented in SOFTSUSY. 

The NSSB 
contributions to the mass matrices given in \cite{JJ}
are the leading, tree-level,
NSSB corrections to the relationship between the physical particle mass and
the running model parameters. SOFTSUSY includes some MSSM loop corrections in
the mass matrices, but such corrections involving the NSSB couplings
have not been added. 

\section{Results on the particle spectrum}
\label{results}

As an example of the superparticle spectrum in NSSB, 
figures~\ref{C+fig} and~\ref{mut+fig} show some of 
the masses on lines through NSSB parameter space passing through mSUGRA
standard point II (defined using the sign conventions of~\cite{ATLASTDR2}):
\begin{eqnarray}
m_0 &=& 400 \GeV; \nonumber \\
M_{1/2} &=& 400 \GeV; \nonumber \\
\tan \beta &=& 10; \nonumber \\
A_0 &=& 0; \nonumber \\
\mu &>& 0.
\label{sugII}
\end{eqnarray}

The same
point with the opposite sign of $\mu$,
has been investigated and shows very
similar results, except with $C_0 \mapsto -C_0$ and $\tilde{\mu} \mapsto
-\tilde{\mu}$.

\FIGURE{
\hbox{\epsfysize=10cm
\epsffile{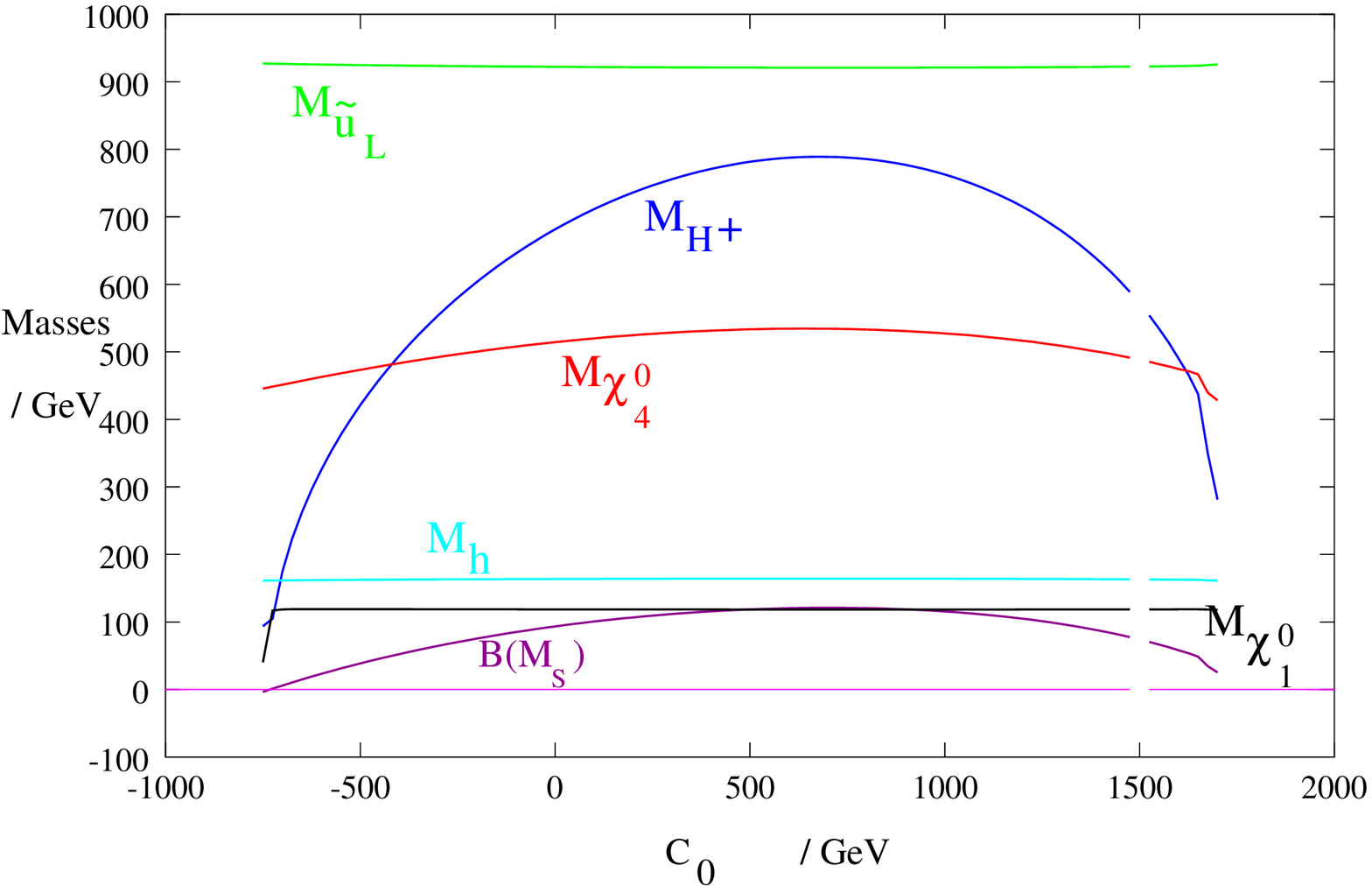}}
\caption{Some of the MSSM particle masses, and the $B$ parameter at the REWSB
scale, for nonzero values of the non-standard supersymmetry breaking
parameter $C_0$. $C_0 = 0$ here is the mSUGRA point II: eq.~(\ref{sugII}). At the missing point,
and the final point, the numerical method failed to converge
correctly. Detail of this region can be seen in figure~\ref{Cextreme}. The
masses and parameters shown are those referred to in
section~\ref{constraints},
or those masses which vary significantly across the range. Those masses shown 
which do not change significantly have been chosen to indicate this fact.
}
\label{C+fig}
}

\FIGURE{
\hbox{\epsfysize=10cm
\epsffile{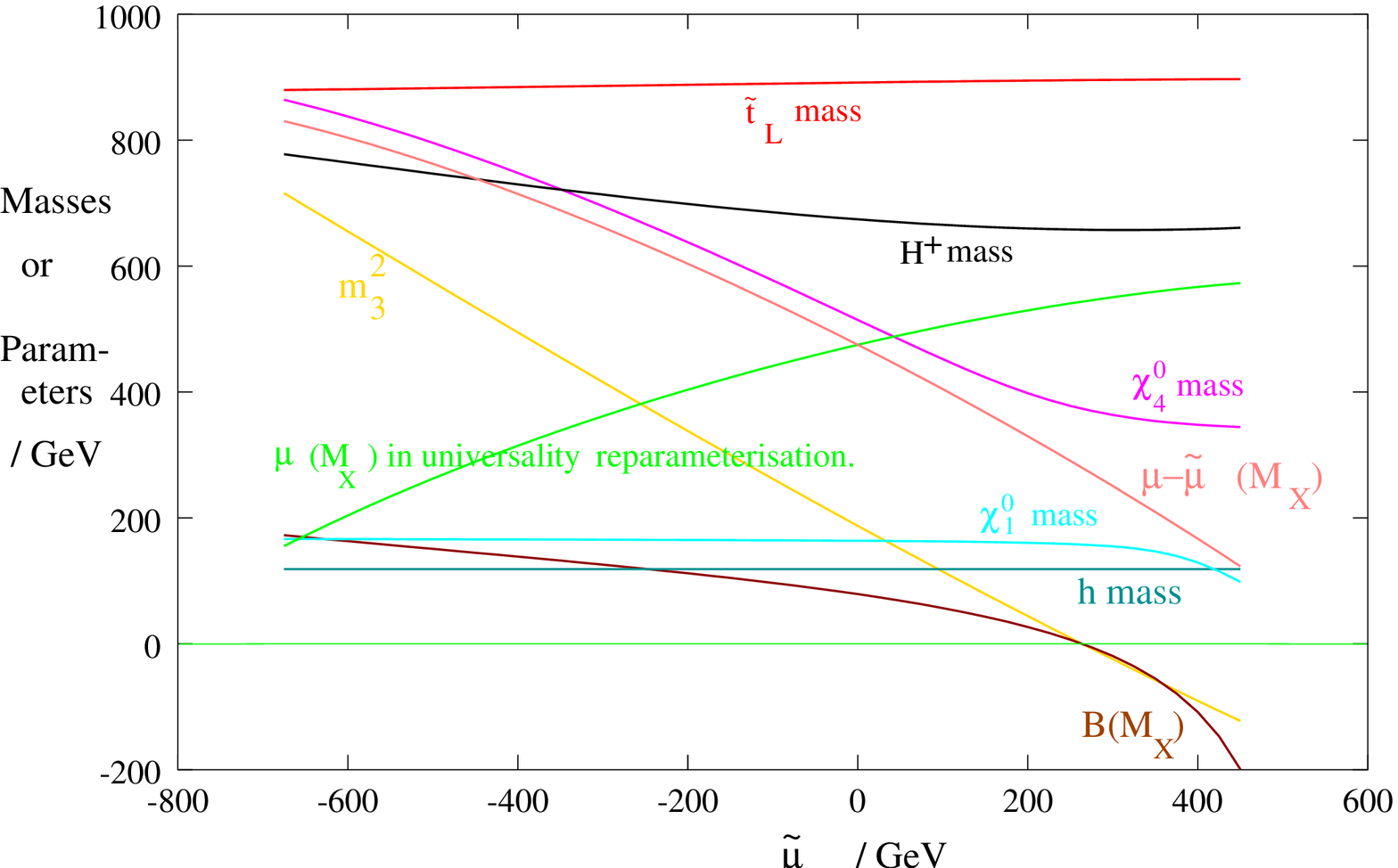}}
\caption{Some of the MSSM parameters and particle masses for nonzero values
of the non-standard supersymmetry breaking parameter
$\tilde{\mu}$. $\tilde{\mu} = 0$ here is the mSUGRA point II: eq.~(\ref{sugII}).
}
\label{mut+fig}
}

Note that there is nothing special about the spectrum at the zero values of
$C_0$ or $\tilde{\mu}$. The MSSM point already has broken supersymmetry. The
NSSB terms break a few remaining equalities between SUSY breaking terms, but
this has no sudden qualitative effect on the superparticle spectrum. 

The masses and soft breakings
for nonzero $C_0$ near mSUGRA point II show an extremum in fig.~\ref{C+fig}.
This occurs where the $B$ parameter is equal at $M_S$ and $M_Z$. 
This behaviour is
a property of the RGEs for the system. 

The very low masses of the normally heavier Higgs bosons ($A, H^\pm, H^0$)
for large negative values of $C_0$, 
as shown in fig.~\ref{C+fig}, and the removal of
all but the lightest neutralinos and charginos from the lower mass part of
the spectrum (fig.~\ref{mut+fig}) for large negative 
values of $\tilde{\mu}$, might
show interestingly different
phenomenology. This may be worth investigating with ISAJET~\cite{ISASUSY} or HERWIG~\cite{HERWIG}.

\section{Parameter space constraints}
\label{constraints}

What determines the largest positive and negative values of $C_0$ and
$\tilde{\mu}$ allowed? For very large negative $C_0$ both the
eigenvalues of the Higgs 1-loop effective potential at $M_S$ become negative
at $\langle H_1 \rangle = \langle H_2 \rangle = 0$. This is in contrast with the usual MSSM case where
only one of the eigenvalues is negative -- a phase transition has occurred. In
this region, the REWSB constraint (eq.~(\ref{REWSBcon}))
fails to correctly place the
minimum of the Higgs effective potential at the required values of $v$ and
$\tan \beta$, placing a saddle point there instead. See fig.~\ref{C+fig}, 
where the Higgs masses become small as we approach the
transition at $C_0\sim -750$~GeV. 
The shape of the surface below the transition is shown in
fig.~\ref{sadplot}. 
Note that eq.~(\ref{REWSBcon})
assumes that the point found does correspond to the
physical minimum,
 so we do {\emph{not}} have a correct determination of the superspectrum in
the unusual phase.

\FIGURE{
\hbox{\epsfysize=8cm
\epsffile{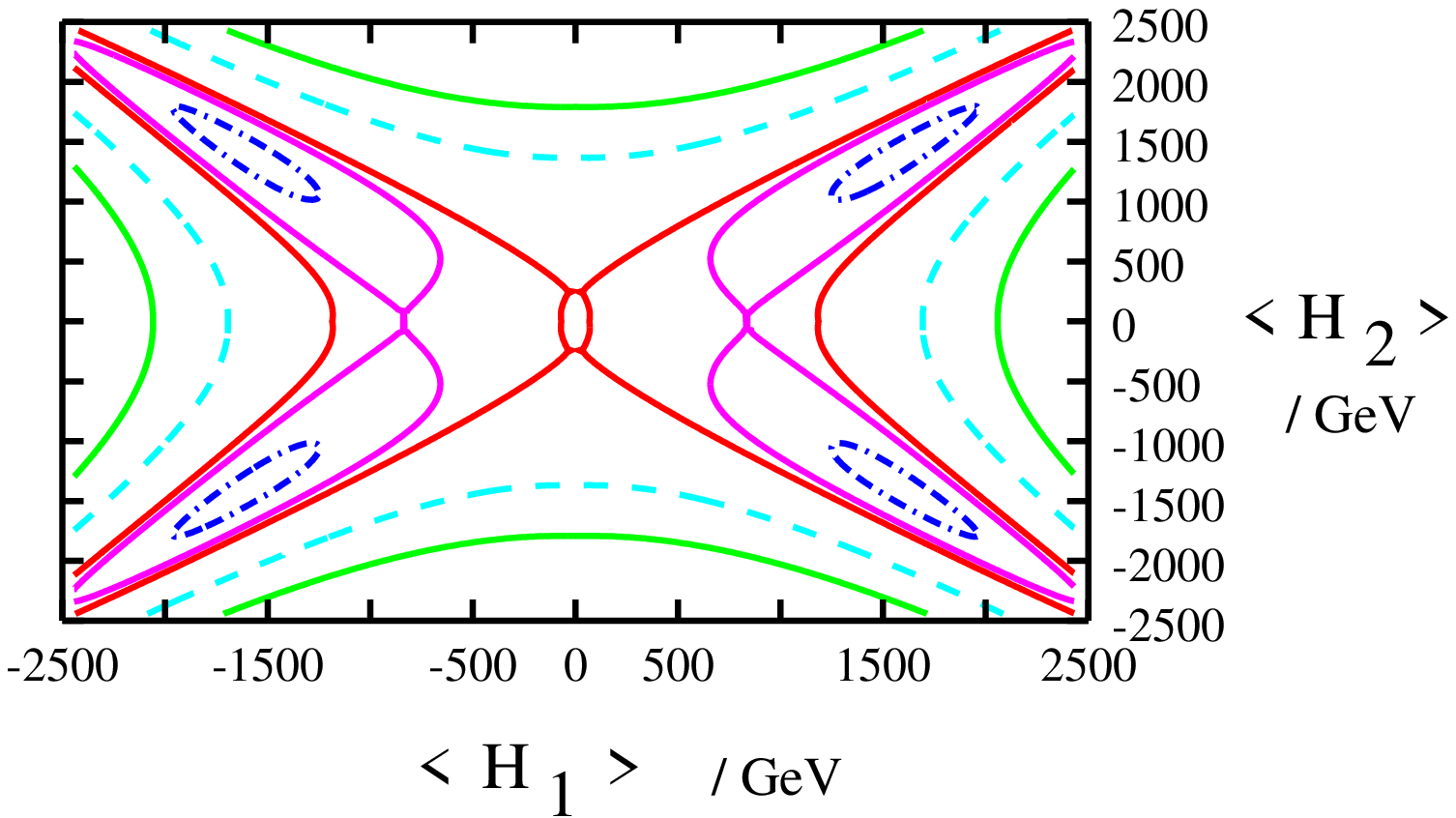}}
\caption{The Higgs effective potential surface for large
values of $C_0$ near
mSUGRA point II. (This plot was made with $C_0 = -800$ GeV.)
Note the saddle point at $\tan \beta = 10, \langle v \rangle
= 243$ \GeV, i.e. at 
$\langle H_2 \rangle \sim 240, \langle H_1 \rangle \sim 0$. The central (red,
solid) contour is $-1 \times 10^8$ GeV$^4$, the next contour (magenta, solid) is at $-8.5
\times 10^9$ GeV$^4$, the small elliptical contours (blue, dot-dashed) are at
$-2 \times 10^{10}$ GeV$^4$, the blue, dashed contour is $10^{11}$ GeV$^4$, and
the outermost (green, solid) contours are at $2 \times 10^{11}$ GeV$^4$.
}
\label{sadplot}
}

In order to obtain the above understanding, and obtain figure~\ref{sadplot}, code was written to obtain the full shape of the Higgs effective potential away from the minimum, including the dominant 1-loop corrections.

We should next attempt to resolve the question: is there a point in the
parameter space where this new phase yields physical values of the particle
masses? For the parameter set shown in fig.~\ref{sadplot} the Higgs VEV is much
too large, resulting in a value of $M_Z$ which is too large. 
Perhaps, for a given pair of $\tan \beta$ and $|\langle H \rangle |$, 
more than one set of
values for the soft breakings is possible, corresponding to each of the two
phases. Although, at tree level, eq.~(\ref{REWSBcon})
suggests a unique value of
the parameters for a given $\tan \beta$ and $|\langle H \rangle|$, it must actually be
solved iteratively when the loop corrections are taken into account, because
these corrections depend on the soft SUSY breakings. 
It is therefore possible that this
iterative procedure possesses more than one solution. However, after an
extensive search of parameter space, nowhere was the new phase found to be
relevant to real physics.

There exist points in parameter space where the ``universality
reparameterisation'' discussed in section~\ref{Lagrangian}
does not exist after imposing the low-energy constraints
and running up to the high scale: the required value of $\mu^2$ in
the universality reparameterisation at high scale becomes negative. 
At such points the Higgs soft masses cannot, for positive $\mu^2$ be made equal to the sfermion soft masses,
while maintaining consistency with the electroweak symmetry breaking
conditions. If this equality is forced, there is no positive
 value of $\mu^2$ for
which electroweak symmetry breaking occurs. This provides the limit on large
negative 
values of $\tilde{\mu}$ at mSUGRA point II.

\FIGURE{
\hbox{\epsfysize=10cm
\epsffile{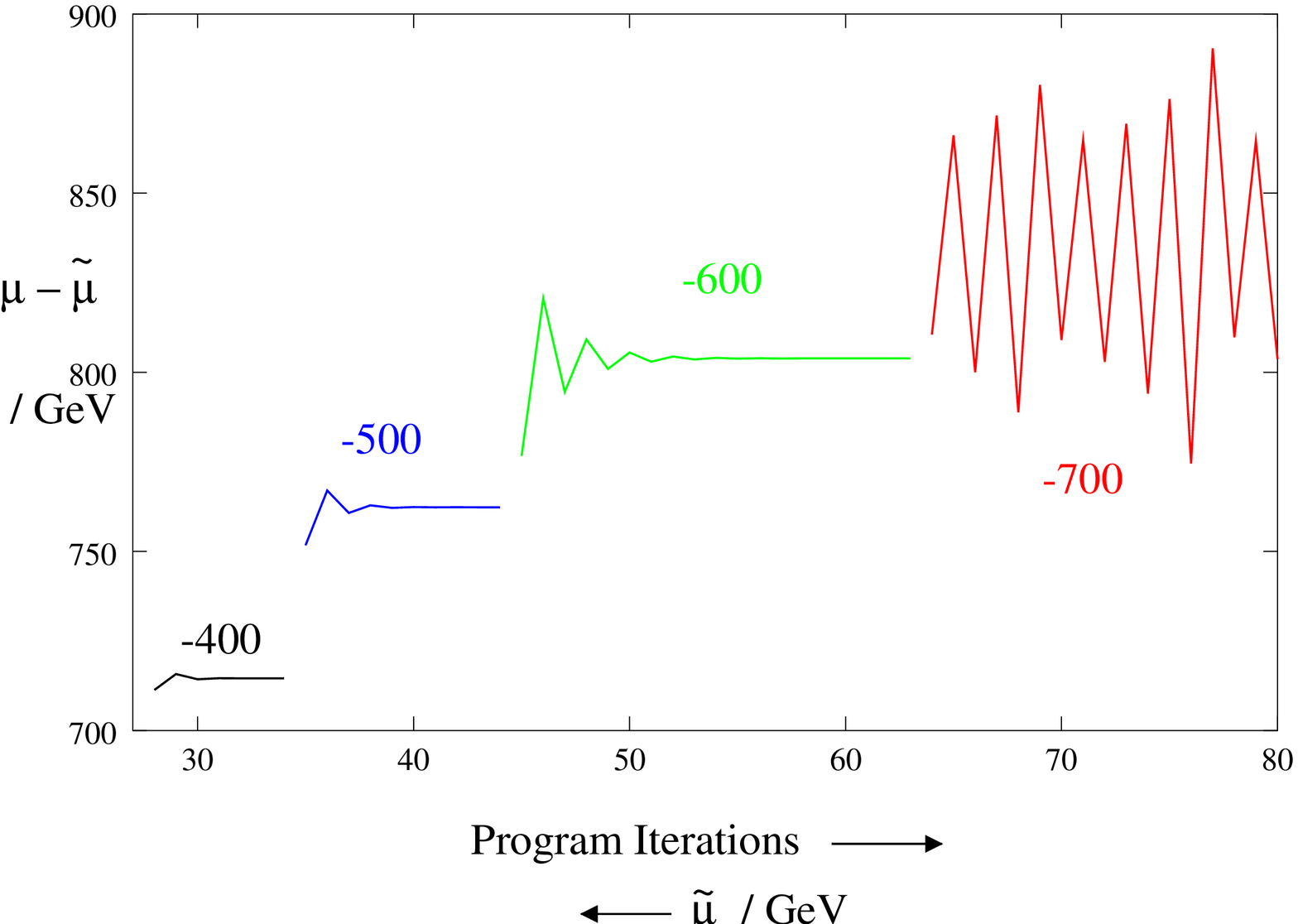}}
\caption{At mSUGRA point II, 
the recurrence relation solution to the SUSY RGEs converges more and
more slowly for increasingly negative values of $\tilde{\mu}$. Each separate
convergence in 
the plot is for a different value of $\tilde{\mu}$, with $\tilde{\mu}$
becoming more negative to the right. Note that for sufficiently large 
negative values of $\tilde{\mu}$,
the method fails to converge at all. The numbers near the lines show the
value of $\tilde{\mu}$ used for that point, in GeV.
}
\label{muvalues}
}

It is for these large negative values of $\tilde{\mu}$ at mSUGRA point II
that the recurrence
relation approach becomes unstable. As shown in figure~\ref{muvalues}, the 
convergence toward the fixed point becomes steadily slower, before it
finally becomes unstable. A naive belief in the recurrence relation approach
would result in an underestimate of the size of the valid parameter space.
The
recurrence relation approach becomes unstable at $\tilde{\mu} \sim -675$~GeV, 
while the
single-ended drift method explained in appendix~\ref{numapp} extends the
region which can be explored as far as $\tilde{\mu} \sim -775$~GeV.
Even that method cannot reach the physical limit, due to
numerical instabilities caused by early iterations
 entering the unphysical region.
This is illustrated in fig.~\ref{muextreme}.

\FIGURE{
\hbox{\epsfysize=10cm
\epsffile{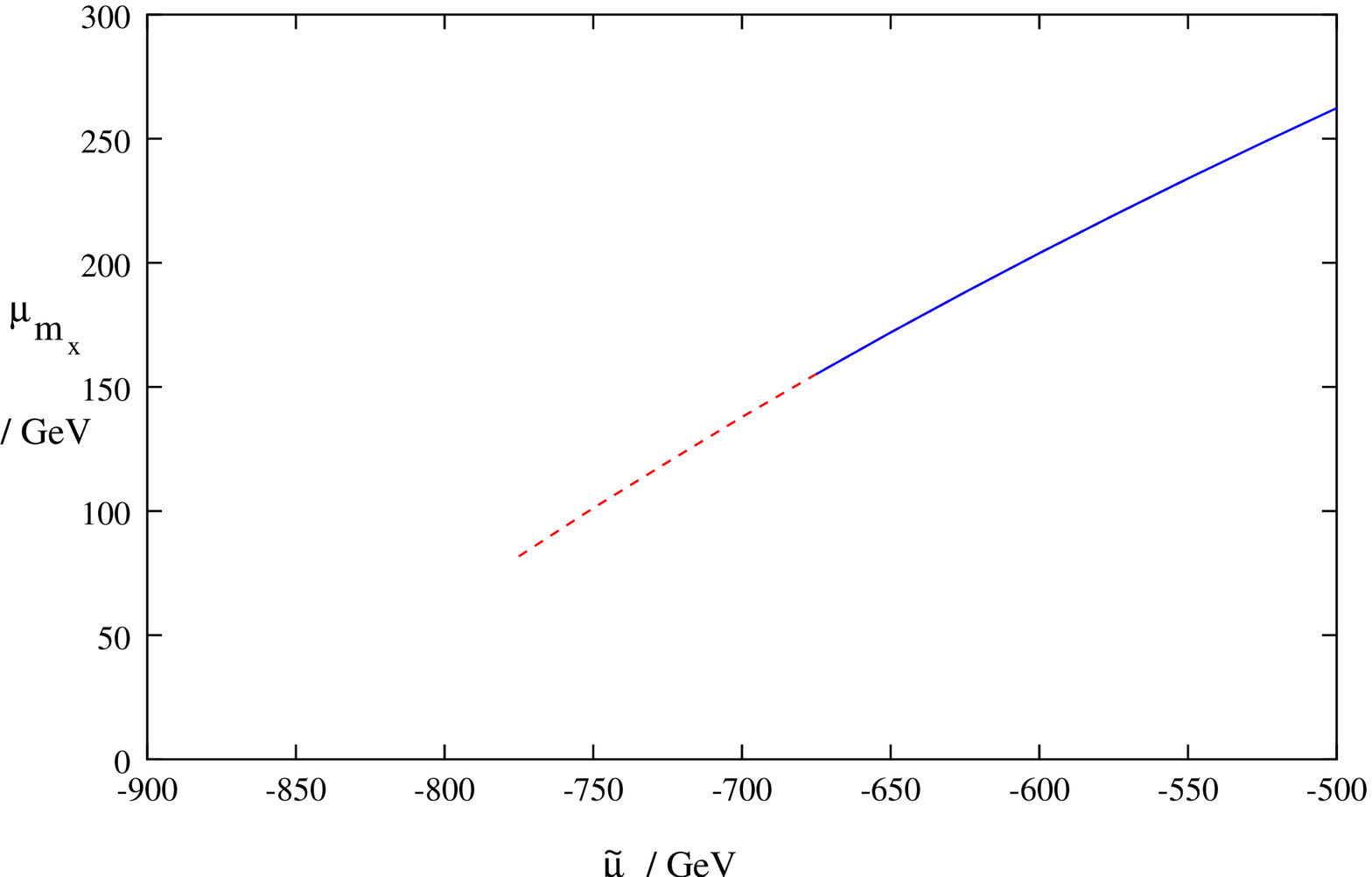}}
\caption{The value of $\mu$ in universality reparameterisation at $M_X$ for
non-zero $\tilde{\mu}$ near mSUGRA point II. The dotted part of the 
line shows the extra reach provided by the single-ended drift method.
}
\label{muextreme}
}

For large positive $\tilde{\mu}$ at mSUGRA point II, the bound on 
$\tilde{\mu}$ due to the value of $B$ runs to extremely large negative values
at high scale, making the calculation unstable. This is an artificial
consequence of using $B$ rather than $m_3^2$. As shown in fig.~\ref{mut+fig},
$m_3^2 = B (\mu - \tilde{\mu}) $ is finite, 
but $\mu - \tilde{\mu}$ is becoming small, so $B$ is
singular. The calculation becomes unstable here due to the large value of $B$,
but the upper bound on the value of $\tilde{\mu}$ comes from the lightest 
neutralino mass, which is becoming small as $(\mu - \tilde{\mu})$ does.

For large positive
values of $C_0$, above around $C_0 \sim 1500$~GeV, the drift method becomes
non-convergent. 
The single-ended shooting method discussed in section~\ref{software} and
appendix~\ref{numapp}
 takes us further, and is used to produce figure~\ref{C+fig}, which takes us
up to $C_0 \sim 1675$~GeV. Extrapolating,
it appears that the physical limit occurs for the same reason as for 
large negative $C_0$, i.e. the Higgs masses become small. However, the
physical limit cannot be reached even with the single-ended shooting method,
for the reason discussed above. With care, some useful data can be teased out
of this region (see fig.~\ref{Cextreme}), showing that the physical 
limit is around $C_0 \sim 1675$~GeV. 

\FIGURE{
\hbox{\epsfysize=10cm
\epsffile{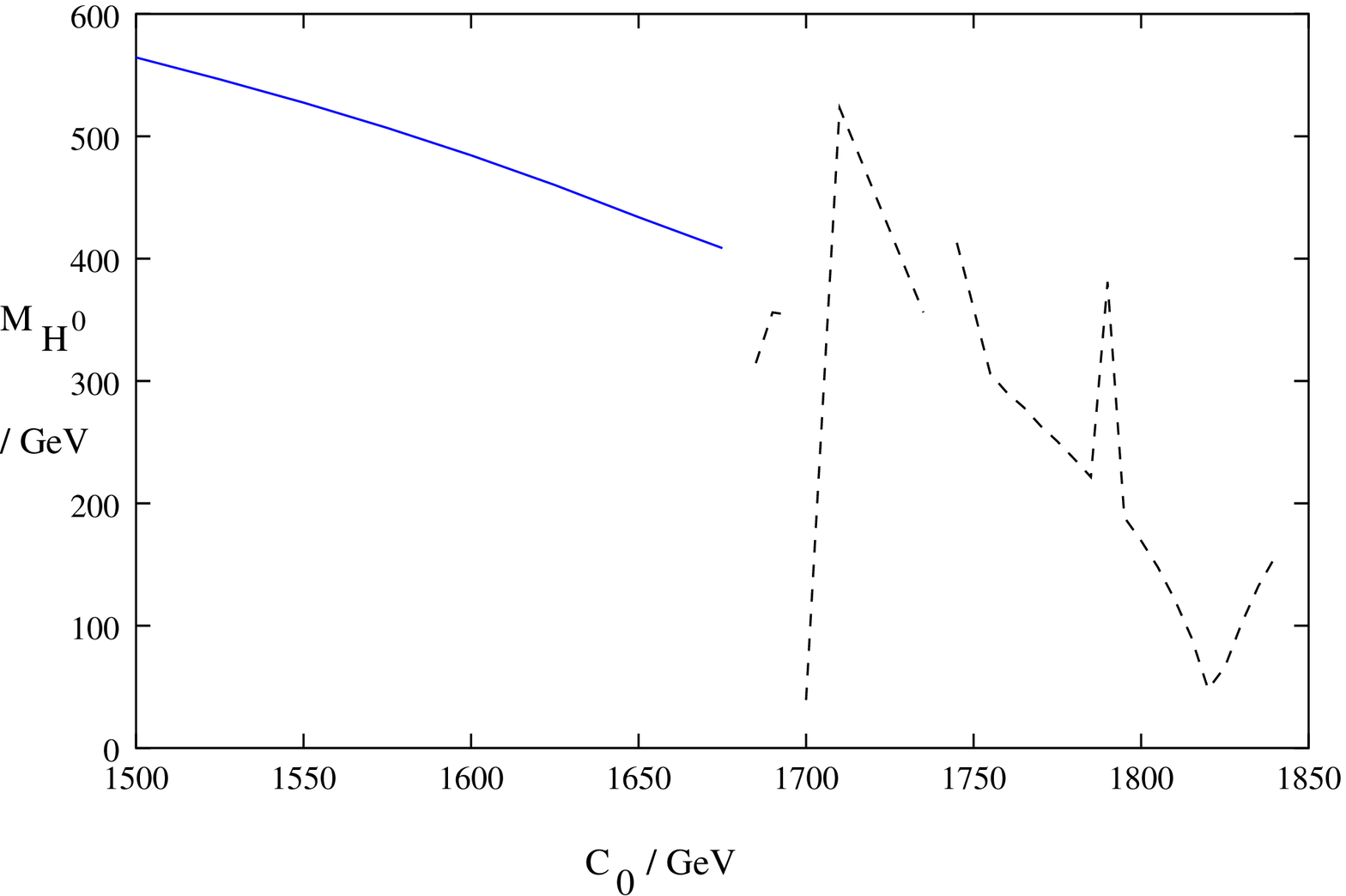}}
\caption{The mass of the heavy neutral Higgs particle,
for non-zero $C_0$ near mSUGRA point II. All the data was produced using the
single-ended shooting method.
The dashed part of the line is the region near the physical limit
where transients spoil the stability of the
method. The figure shows 
the outcome of the method, whether or not it has formally
converged. 
}
\label{Cextreme}
}

\section{Fine-tuning}
\label{ft}

Eq.~(\ref{REWSBcon}) shows that, with $m_{H_1}$ and $m_{H_2}$ of order
the mass scale of the scalar superparticles, 
a cancellation is required so that $M_Z$ can take its known value. The larger
the typical mass of the scalar superparticles (controlled by the mSUGRA
parameter $m_0$) the more finely tuned this calculation must be. One possible
numerical measure of the delicacy of this calculation 
is~\cite{ftreach,focuspoints}: 
\begin{equation}
c_a \equiv \left| \frac{\partial \ln M_Z^2}{\partial \ln |a|} \right|.
\end{equation}
The fine-tunings with respect to each model parameter $a$ are calculated, and
the overall fine-tuning of the model is defined to be $c=\mbox{max}(c_a)$.

Here, there is a subtlety, again caused by the presence of the
reparameterisation invariance: 
the same Lagrangian may be
described by a different choice of parameters. For example, one could choose,
instead of $C_0$ and $\tilde{\mu}$, to reparameterise so that $C_0 = 0$, and
use $(\overline{m_{\tilde{q},\tilde{l}}} - \overline{m_h})$ as
the other parameter. Then, one would get different fine tuning values
for the same Lagrangian, depending on ones choice of parameter set. If
this situation is thought to be unsatisfactory, the solution is to take the
fine tuning with respect to the reparameterisation-invariant combinations of
parameters appearing in the Lagrangian, for example, $(\tilde{\mu} - \mu)$.

On the other hand, one may view it as acceptable that the fine-tuning depends
upon the choice of reparameterisation, given that the fine-tuning measure is
supposed to be a measure of ones satisfaction with the deeper model lying
beyond the Lagrangian being studied. A parameterisation choice reveals
something of ones assumptions about the beyond-the-MSSM physics giving rise
to the MSSM-type Lagrangian.

Fig.~\ref{ftplot1} shows fine-tuning contours near
mSUGRA point II with respect to invariant combinations of parameters,
and fig.~\ref{ftplot2} for the parameter set chosen here.

\FIGURE{
\hbox{\epsfysize=7cm
\epsffile{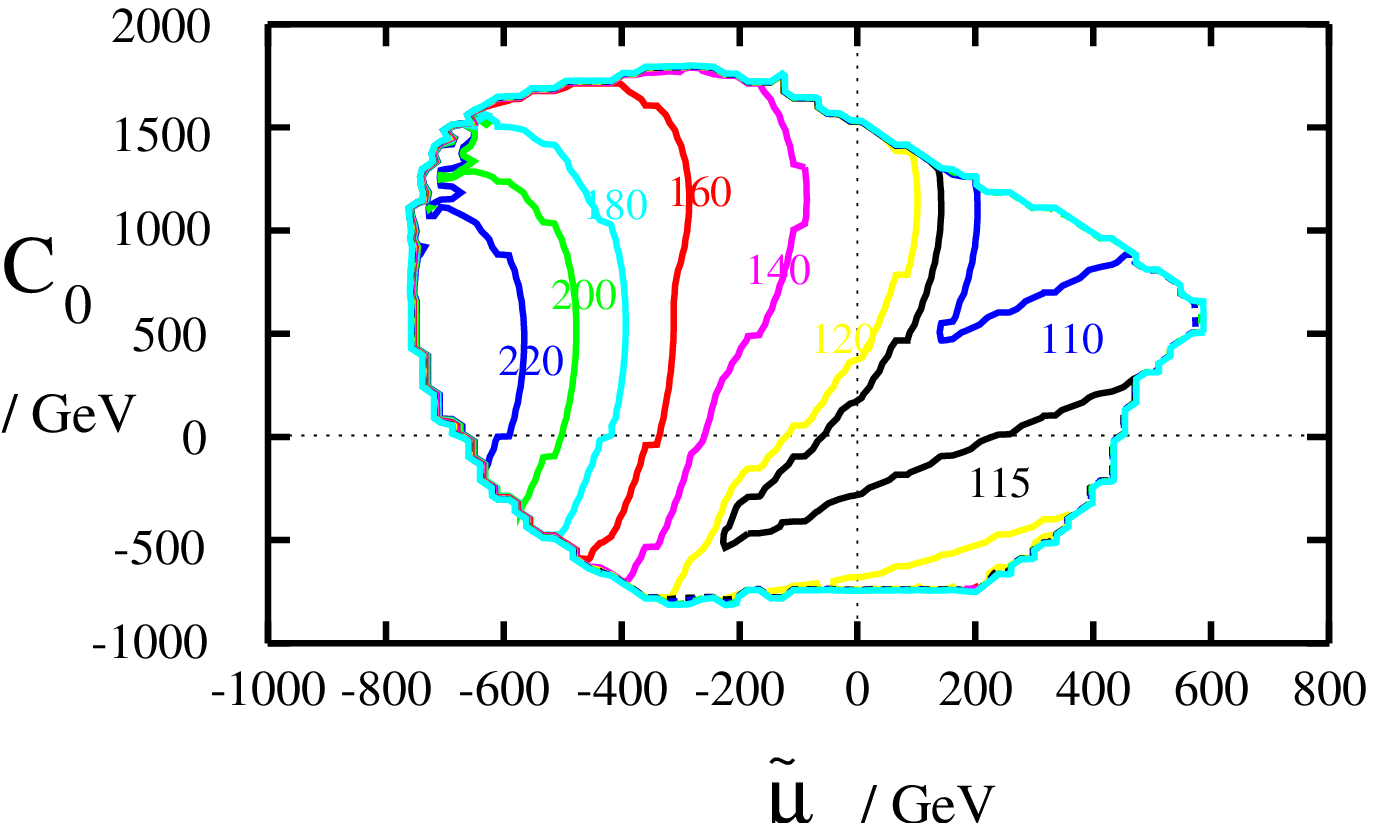}}
\caption{Fine tuning contours for non-standard supersymmetry breaking, with
fine-tuning differentials taken with respect to invariant combinations of
parameters. The plane is shown parameterised in terms of $C_0$ and
$\tilde{\mu}$. The outermost
line shows the edge of the valid parameter space, 
according to the drift method.
}
\label{ftplot1}
}

\FIGURE{
\hbox{\epsfysize=7cm
\epsffile{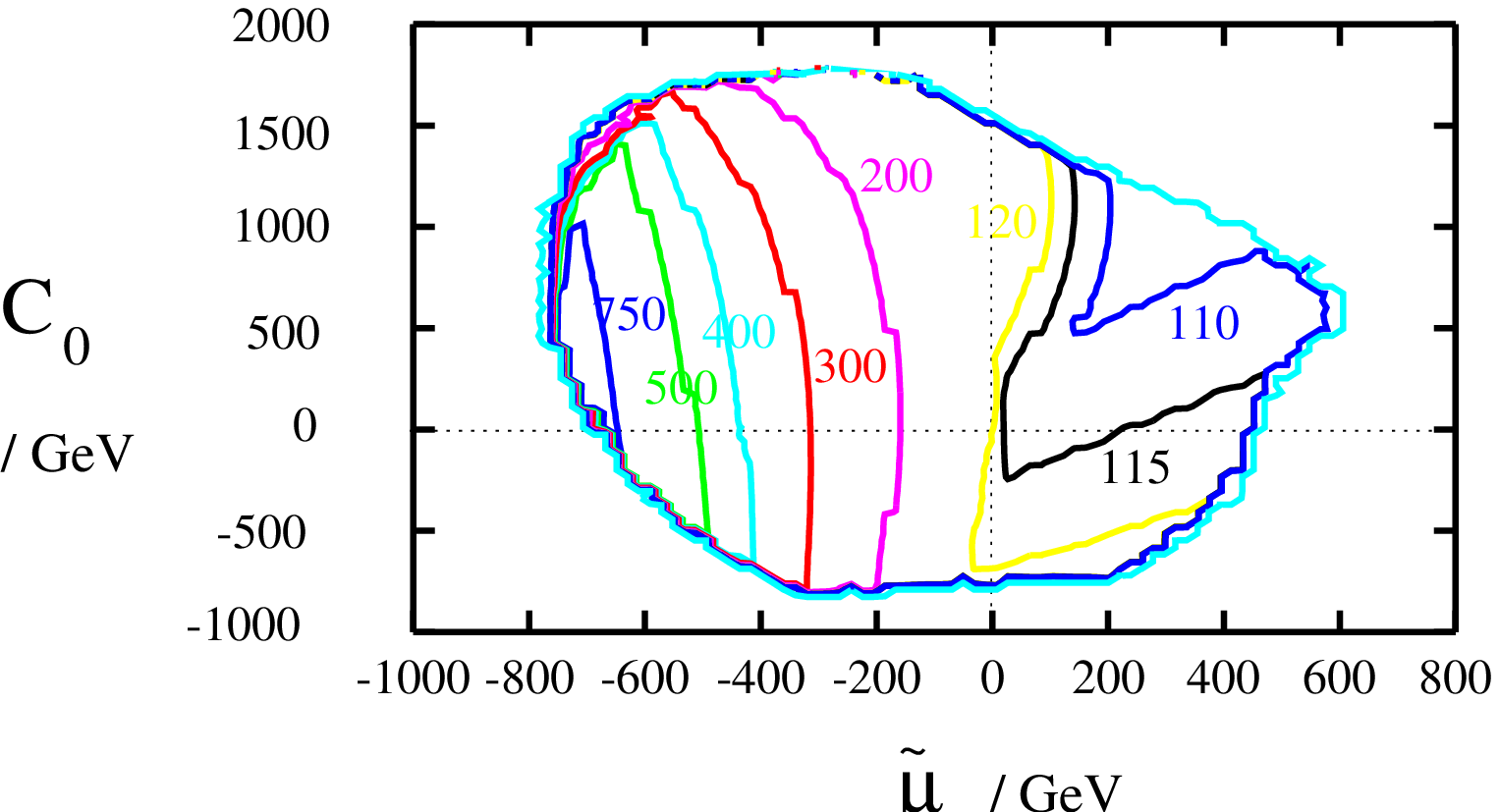}}
\caption{Fine tuning contours for non-standard supersymmetry breaking, with
fine-tuning differentials taken with respect to the usual mSUGRA parameters,
$C_0$ and $\tilde{\mu}$.
}
\label{ftplot2}
}

These results show that, in both cases, 
there is significant fine-tuning
with respect to negative $\tilde{\mu}$. There is no significant fine-tuning
with respect to $C_0$, and for positive  $\tilde{\mu}$ the dominant fine-tuning
is that with respect to the standard supersymmetry breakings. These effects
are consistent with the behaviour explained in section~\ref{constraints}.
The absolute magnitude of the fine tuning is smaller when
considered with respect to the invariant combinations of parameters.

\section{Summary}
\label{summary}	

The 3-family beta functions have been obtained for the MSSM with the NSSB
terms, in a form suitable for numerical implementation, and in an economical
way making use of the reparameterisation invariance.

The mSUGRA boundary conditions have been adapted for use in the context of
the NSSB reparameterisation invariance, by choosing to impose the boundary
conditions in the reparameterisation where they are most nearly satisfied.
These boundary conditions allow the
conventional mSUGRA (or CMSSM) model to be seen as a subspace of the full
parameter space including NSSB terms, and the calculation of
the superspectrum can be made in the conventional way.

The superparticle spectrum and a fine-tuning measure have been calculated
for non-standard SUSY breaking near mSUGRA point II, where it is found that
there is no significant experimental signature indicating the new terms,
except where they approach their maximum allowed values.

The conventional MSSM methods are, however, found to be 
inadequate in this extreme region, due to the instability of the numerical
method used. A more stable numerical method
has been developed which solves this problem. 

The phase transition in the Higgs potential occurring for large negative $C_0$
at mSUGRA point II suggests that this model potentially has a rich phase
structure, although this appears to be of only theoretical significance,
since in this phase, the Higgs VEVs are very large.

In conclusion, we see that the numerical approach used to calculate the MSSM
superspectrum can be applied to this model, but there are subtle complexities
which must be dealt with. As supersymmetric models continue to be developed,
given the need to calculate their properties 
sufficiently accurately for comparison with experiment, 
such subtleties must be treated carefully.

\acknowledgments

I would like to thank 
the members of the Cambridge Supersymmetry Working Group, particularly
B.C. Allanach, the author of SOFTSUSY, and B.R. Webber, my supervisor.
This work was funded by the U.K. Particle Physics and
Astronomy Research Council.

\pagebreak

\appendix

\section{Renormalisation group equations}
\label{RGEapp}

The 1-loop three family $\beta$-functions for the MSSM 
are presented below, in the
reparameterisation where $\mu = 0$, 
with family, colour and electroweak indices suppressed. The convention for
the $\beta$-functions of ref.~\cite{barger} is used, i.e.:
\begin{equation}
\beta_x = \frac{dx}{d(\ln m)},
\end{equation}
where $m$ is the renormalisation scale.

Note that for the $\beta$-functions of the scalar masses,
only the additional contribution
from the non-standard terms is given. These should be added to the
expressions given in~\cite{barger}, which is denoted as
$\beta^{MSSM}$. The $\beta^{MSSM}$ in SOFTSUSY include some two-loop
terms. For $\beta$-functions not given below, the NSSB couplings do not
contribute, and the results of~\cite{barger} may be used.

\def\Tr{\mbox{Tr}}
\begin{eqnarray}
16 \pi^2 \beta_{C^u} & = & 3 \Tr(Y^d Y^{d \dag}) C^u + \Tr(Y^e
 Y^{e \dag}) C^u + 2 C^u Y^{u \dag} Y^u + Y^u
 Y^{u \dag} C^u \nonumber \\
&& + 6 Y^u \Tr (C^u Y^{u \dag})
 + 3 Y^d Y^{d \dag} C^u + 2 C^d Y^{d \dag} Y^u , \nonumber \\
&& - C^u(\frac{4}{9} g_1^2 + \frac{16}{3} g_3^2) - 4 \tilde{\mu} Y^d Y^{d \dag}
 Y^u - 2 \tilde{\mu} Y^u(g_1^2 + 3 g_3^2), \\
16 \pi^2 \beta_{C^d} & = & 3 \Tr(Y^u Y^{u \dag}) C^d 
 + 2 C^d Y^{d \dag} Y^d + Y^d
 Y^{d \dag} C^d  \nonumber \\
&& + 6 Y^d \Tr (C^d Y^{d \dag})
 + 3 Y^u Y^{u \dag} C^d + 2 C^u Y^{u \dag} Y^d 
	+ 2 Y^d \Tr (C^e Y^{e \dag}) , \nonumber \\
&& + C^d(\frac{2}{9} g_1^2 - \frac{16}{3} g_3^2) 
- 4 \tilde{\mu} Y^u Y^{u \dag} Y^d - 2 \tilde{\mu} Y^d(g_1^2 + 3 g_3^2), \\
16 \pi^2 \beta_{C^e} & = & 3 \Tr(Y^u Y^{u \dag}) C^e + 2 C^e Y^{e \dag}
 Y^e + Y^e Y^{e \dag} C^e  \nonumber \\
&& + 2 Y^e \Tr (C^e Y^{e \dag})
 + 6 Y^e \Tr(C^d Y^{e \dag})  \nonumber \\
&& - 2 C^e g_1^2
 - 2 \tilde{\mu} Y^e(g_1^2 + 3 g_3^2) \\
16 \pi^2 \beta_{m^2_{H_1}} & = &16 \pi^2 \beta^{MSSM}_{m^2_{H_1}}
	+ 6 \Tr (C^u C^{u \dag}) - (6 g_2^2 + 2 g_1^2) \tilde{\mu}^2, \\
16 \pi^2 \beta_{m^2_{H_2}} & = & 16 \pi^2 \beta^{MSSM}_{m^2_{H_2}}
	+ 6 \Tr (C^d C^{d \dag}) + 2 \Tr(C^e C^{e \dag})
	- (6 g_2^2 + 2 g_1^2) \tilde{\mu}^2, \\
16 \pi^2 \beta_{m^2_{\tilde{q}_L}} 
	&=& 16 \pi^2 \beta^{MSSM}_{m^2_{\tilde{q}_L}}
+ 2 (C^u C^{u \dag} + C^d C^{d \dag})
 - 4\tilde{\mu}^2 (Y^u Y^{u \dag} + Y^d Y^{d \dag}), \\
16 \pi^2 \beta_{m^2_{\tilde{l}_L}} 
	&=& 16 \pi^2 \beta^{MSSM}_{m^2_{\tilde{l}_L}}
	+ 2 C^e C^{e \dag}
 - 4\tilde{\mu}^2 Y^e Y^{e \dag}, \\
16 \pi^2 \beta_{m^2_{\tilde{u}_R}} 
	&=& 16 \pi^2 \beta^{MSSM}_{m^2_{\tilde{u}_R}}
	+ 4 C^{u \dag} C^u
	 - 8\tilde{\mu}^2 Y^{u \dag} Y^u, \\
16 \pi^2 \beta_{m^2_{\tilde{d}_R}} 
	&=& 16 \pi^2 \beta^{MSSM}_{m^2_{\tilde{d}_R}}
	+ 4 C^{d \dag} C^d
	 - 8\tilde{\mu}^2 Y^{d \dag} Y^d, \\
16 \pi^2 \beta_{m^2_{\tilde{e}_R}} 
	&=& 16 \pi^2 \beta^{MSSM}_{m^2_{\tilde{e}_R}}
	+ 4 C^{e \dag} C^e
	 - 8\tilde{\mu}^2 Y^{e \dag} Y^e, \\
16 \pi^2 \beta_B &=& \frac{2}{\tilde{\mu}} \Tr(3 C^{u \dag} A_u + 3 C^{d \dag}
A_d + C^{e \dag} A_e ) + 2 g_1^2 M_1 + 6 g_2^2 M_2 ,\\
16 \pi^2 \beta_{\tilde{\mu}} &=&
 \tilde{\mu} \Tr(3 Y^u Y^{u \dag} + 3 Y^d Y^{d \dag} + Y^e Y^{e \dag}) -
g_1^2 - 3 g_2^2.
\end{eqnarray}

\section{Numerical methods}
\label{numapp}

\subsection{Introduction}

The numerical methods used in this paper
may be formally described 
as follows, where for clarity we note what each general variable corresponds
to in our SUSY case. Let the boundary conditions at scale $x=M_X$ be 
\begin{equation}
f(p(x)) = 0.
\label{BC}
\end{equation}
This equation corresponds to the mSUGRA boundary conditions, e.g. $m^2_q =
m^2_0$ and $p(x)$ corresponds to
$m_0$, $M_{1/2}$, etc.
Let the boundary conditions at scale $z$ ($=M_Z$ and/or $M_S$) be 
\begin{equation}
g(q(z))=0.
\label{BC2}
\end{equation}
This condition corresponds to the physical values of the
Yukawa couplings, ($h_t$ etc.), the gauge couplings,
and the REWSB condition (which constrains
$\mu$ and $B$). The full set of parameters for the 
model is $\{p(t),q(t)\}$, with $p(t)$ constrained at $t=x$ and $q(t)$ constrained at
$t=z$. Let $q_C$ be the $q$ solving eq.~(\ref{BC2}) and let $p_C$ be the $p$
which solves eq.~(\ref{BC}). The differential equations (renormalisation group equations) then relate the parameters at the two scales:
\begin{eqnarray}
q(z) = R^q(q(x),p(x)) \nonumber \\
p(z) = R^p(q(x),p(x)).
\label{RGEshoot}
\end{eqnarray} 
\subsection{The shooting method}
For a given $q(x)$, i.e. $\mu, B , Y^f \tan \beta$, and $g_{1,2,3}$,  at $M_X$,
therefore, $g(q(z))$
can be calculated numerically through $R^q$. The solution of the boundary value
problem therefore reduces to the numerical solution of the algebraic equation
$g(R^q(q(x),p_C))=0$ for $q(x)$. This is the ``shooting method'' of~\cite{numrec}.

The shooting method cannot, however, be used for SUSY.
Firstly, we have three scales at
which boundary conditions are imposed: $M_Z, M_S$ and $M_X$. Secondly, the
boundary conditions at $M_Z$ and $M_S$ cannot be conveniently expressed in
the form of eq.~(\ref{BC2}).

\subsection{The drift method}

In most implementations of the shooting method for SUSY, an alternative
technique, which can be called the ``drift method'',
is used: Start with a guess, $q^g$, for $q(x)$, i.e. $\mu(M_X)$ etc,
and determine
$p(x)$ from eq.~(\ref{BC}). Determine $q(z)$ and $p(z)$ from
eq.~(\ref{RGEshoot}), the renormalisation group equations.
Impose eq.~(\ref{BC2}) to re-determine $q(z)$. (Set the Yukawa matrices and
gauge couplings from experimental data
and set $\mu$ and $B$ from the REWSB constraints.) Use the inversion
of~\ref{RGEshoot}, i.e. run back up to high scale, to find new values for
$q(x)$ and $p(x)$, $q'(x)$ and $p'(x)$ respectively. 
This corresponds to a recurrence relation $q'(x) = Q(q^g)$. Alternatively,
the same procedure can be applied with a guess $p^g$ for $p(z)$, i.e., the soft
breaking parameters at $M_Z$, and a recurrence
relation for $p(z)$.

The required values for $p,q$ consistent with both sets of boundary
conditions will be a fixed point of $Q$. We can hopefully determine this
fixed point by making a guess for $q(x)$ and iterating the procedure until it
is convergent, i.e. hoping the recurrence relation will drift to the fixed
point we want. Convergence is tested for by requiring $q$ and $p$ not to
change over a cycle.
However, note that the fixed point corresponding to the
physical solution is not guaranteed to be the only fixed point, nor is it
guaranteed to be stable.

\subsection{The single ended shooting method}

When the required fixed point of Q becomes unstable, and/or 
the drift method drifts
into a different fixed point of Q, a different approach can be used.
The equation:
\begin{equation}
Q(q^g)-q^g = 0 \label{sol}
\end{equation}
can be solved for $q^g$. In this work, the Newton-Raphson method was used to
solve this equation. This has improved
stability, but is not perfect, because any unrequired fixed points of $Q$
will also correspond to roots of eq.~(\ref{sol}). The method can be slow,
but, if an approximate solution only is desired, one can just take initial
guesses for the
less important parts of $q^g$ such as $Y^e$, rather than solving for
them, making the method faster by reducing the parameter space to be solved.

The procedure used to generate $q'$ also generates
$p' = P(q^g)$. For $q^g$ consistent with both sets of boundary
conditions (and therefore a solution of eq.~(\ref{sol})),
$p'=p_C$, and therefore $f(p')=0$.
This provides a check that one has found the desired root.

One might imagine that the boundary value problem can be reduced
to the numerical solution of the algebraic equation $P(q^g)-p_C = 0$, but
this turns out to be unstable: formally, it is overconstrained
(there being more elements in $p$ than in $q$ for SUSY), but many of the
constraint equations are approximately degenerate.

\end{document}